\documentclass[prb,
                    preprint
                    ]{revtex4}

\usepackage{times}
\usepackage{subfigure}
\usepackage{epsfig}

\date{}

\def\_#1{{\bf #1}}

\def\.{\cdot}

\def\l#1{\label{eq:#1}}
\def\r#1{(\ref{eq:#1})}
\def\=#1{\overline{\overline #1}}

\begin{document}




\title{Cloaking a metal object from an electromagnetic pulse:\\ A comparison between various cloaking techniques}

\author{Pekka Alitalo$^{\ast}$, Henrik Kettunen, and Sergei Tretyakov}

\affiliation{Department of Radio Science and Engineering/SMARAD
Center of Excellence,\\ TKK Helsinki
University of Technology,\\ P.O. Box 3000, FI-02015 TKK, Finland\\
\\
\normalsize{$^\ast$To whom correspondence should be addressed,
E-mail: pekka.alitalo@tkk.fi}\\ }

\date{\today}

\begin{abstract}
Electromagnetic cloaks are devices that can be used to reduce the
total scattering cross section of various objects. An ideal cloak
removes all scattering from an object and thus makes this object
``invisible'' to the electromagnetic fields that impinge on the
object. However, ideal cloaking appears to be possible only at a
single frequency. To study cloaking from an electromagnetic pulse
we consider propagation of a pulse inside a waveguide with a
cloaked metal object inside. There are several ways to achieve
cloaking and in this paper we study three such methods, namely,
the coordinate-transformation cloak, the transmission-line cloak,
and the metal-plate cloak. In the case of the two last cloaks,
pulse propagation is studied using experimental data whereas the
coordinate-transformation cloak is studied with numerical
simulations. The results show that, at least in the studied cases
where the cloaked object's diameter is smaller than the
wavelength, the cloaks based on transmission-line meshes and metal
plates have wider bandwidth than the coordinate-transformation
cloak.
\end{abstract}

\maketitle

\section{Introduction}

The possibility of electromagnetic cloaking, or, in other words,
the reduction of an object's total scattering cross section, has
intrigued researchers for a long time. Already in the 1960s and
1970s a few publications considered this
possibility.\cite{Dolin,Kerker} But it was not until a few years
ago when the topic became widely studied among the
electromagnetics and physics communities after the publication of
papers [3--6]. Several methods can be used to achieve
electromagnetic cloaking, all of them having their own benefits
and drawbacks as can be concluded from recently published review
papers on this topic.\cite{Alu_review,Alitalo_review} As has been
previously concluded,\cite{Alitalo_review} some cloaking methods
suffer from inherently strong dispersion that will unavoidably not
only limit the cloaking bandwidth but also distort a pulse when it
travels through the cloak. In this paper we make a comparative
study of cloaking of a metal object from an electromagnetic pulse.

We study three cloaking methods: The coordinate-transformation
(CT) cloak,\cite{Pendry2006,Schurig2006} the transmission-line
(TL)
cloak,\cite{Alitalo_TLcloakTAP,Alitalo_TLcloakAPL,Alitalo_TLcloakAPS2009,Alitalo_TLcloakMOTL}
and the metal-plate (MP)
cloak.\cite{Alitalo_PPWGcloakMETA,Tretyakov_PPWGcloakPRL} Due to
the moderate dispersion in the TL and MP cloaks, we expect that
the cloaking bandwidth is wider and pulse distortion is mitigated
in these cloaks, as compared to the CT cloak. The cloaked object
is limited in shape when using the TL cloak (a transmission-line
network must fit inside the cloaked object) whereas the MP cloak
and the CT cloak can be used to ``hide'' bulk objects, such as
solid metal cylinders in this case.

For the study of pulse propagation in case of the TL and MP cloaks
we use transmission data measured in a waveguide with a cloak and
a cloaked object
inside.\cite{Alitalo_TLcloakAPL,Tretyakov_PPWGcloakPRL} In case of
the CT cloak, we rely on numerical simulations of an idealized
cloak having a radially varying permeability\cite{Schurig2006}
$\mu$ and inserted inside a similar waveguide as the other cloaks.
To have physically sound values for $\mu$ as a function of the
frequency, we assume that $\mu$ has the Lorentzian type dispersion
characteristics. Pulse propagation through CT cloaks that are
placed in free space and are impinged on by plane waves has been
studied analytically and numerically in several
papers.\cite{Zhang1,Chen,Zhang2,Argyropoulos} Since in this study
the cloaks are placed inside a waveguide with input and output
ports, the distortion of the pulse travelling through the cloaks
is very easy to illustrate and compare to the case of the empty
waveguide.

\section{Studied cloaking devices}
\label{studied_cloaks}

\subsection{Transmission-line cloak}

The transmission-line cloak consists of layered two-dimensional
networks of parallel-strip transmission
lines.\cite{Alitalo_TLcloakAPL} The cloak's outer diameter is
approximately 71~mm while the cloaked object consists of 21 metal
rods having the diameter 2~mm and the array period 5~mm. The
largest outer dimension of the array in the $xy$-plane is 24.5~mm.
It is obvious that due to the periodicity of the network an
electrically large solid object cannot be cloaked with this
method.

Here we use the measured data obtained with a rectangular
waveguide having a passband around 3~GHz and the TL cloak
enclosing the metal rod array placed inside the
waveguide.\cite{Alitalo_TLcloakAPL} The cloak is made by etching
thin sheets composed of an alloy of bronze and beryllium. See
Fig.~\ref{cloaks}a for an illustration of the measurement setup.
The width and height of the waveguide are 86.36~mm and 36.8~mm,
respectively and the distance between ports 1 and 2 is 389~mm.

Although the cloaked object in this case is a two-dimensional
array of metal rods, it should be noted that with the same cloak
also a more complicated three-dimensional object can be
cloaked.\cite{Alitalo_review,Alitalo_TLcloakAPS2009,Alitalo_TLcloakMOTL}


\subsection{Metal-plate cloak}

The metal-plate cloak consists of layered cylindrical metal plates
that have a circular hole cut inside them (the volume which is
cloaked).\cite{Alitalo_PPWGcloakMETA,Tretyakov_PPWGcloakPRL} Due
to the fact that the height of these waveguide plates decreases
radially as moving from the outer surface of the cloak towards the
inside, the electric field which is polarized orthogonally to the
metal plates travels inside these waveguides around the cloaked
region. At the inner surface of the cloak the fields cannot couple
into the cloaked region since the fields are concentrated in a
very small volume due to the small height of the
waveguides.\cite{Tretyakov_PPWGcloakPRL} Note that this cloak
operates for $z$-polarized electric fields in contrast to a
similar cloak geometry which is based on a different design
concept.\cite{Shalaev}

The inner and outer diameters of the MP cloak are 32~mm and 70~mm,
respectively, whereas the diameter of the cloaked cylinder is
30~mm. The plates of the cloak are made of copper and the cloaked
cylinder is made of steel. The measurement data for the MP cloak
is obtained with the same waveguide as for the TL
cloak.\cite{Tretyakov_PPWGcloakPRL} See Fig.~\ref{cloaks}b for an
illustration of the measurement setup.

\subsection{Coordinate-transformation cloak}

\begin{figure} [h!]
\subfigure[]{\epsfig{file=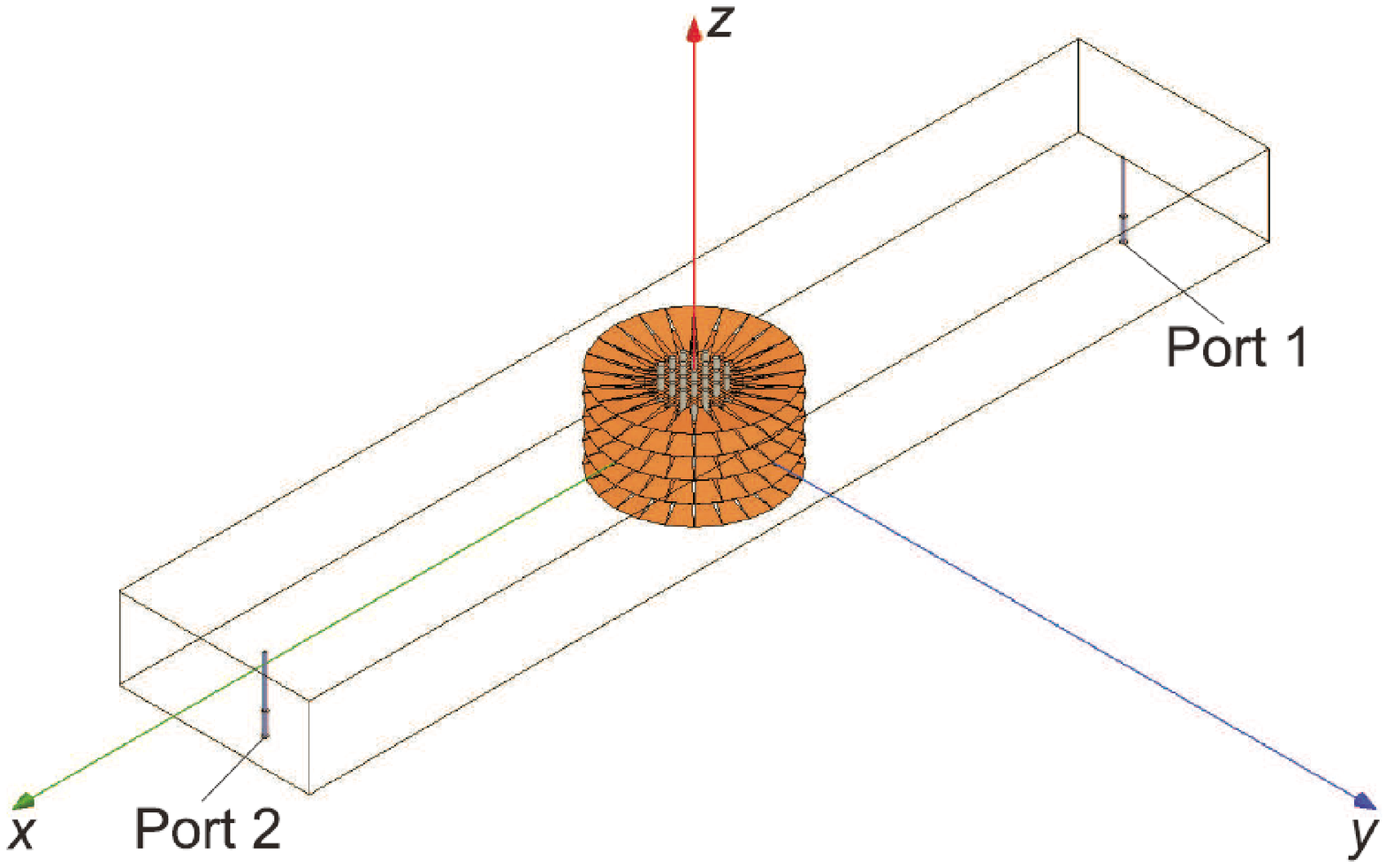, width=0.45\textwidth}}
\subfigure[]{\epsfig{file=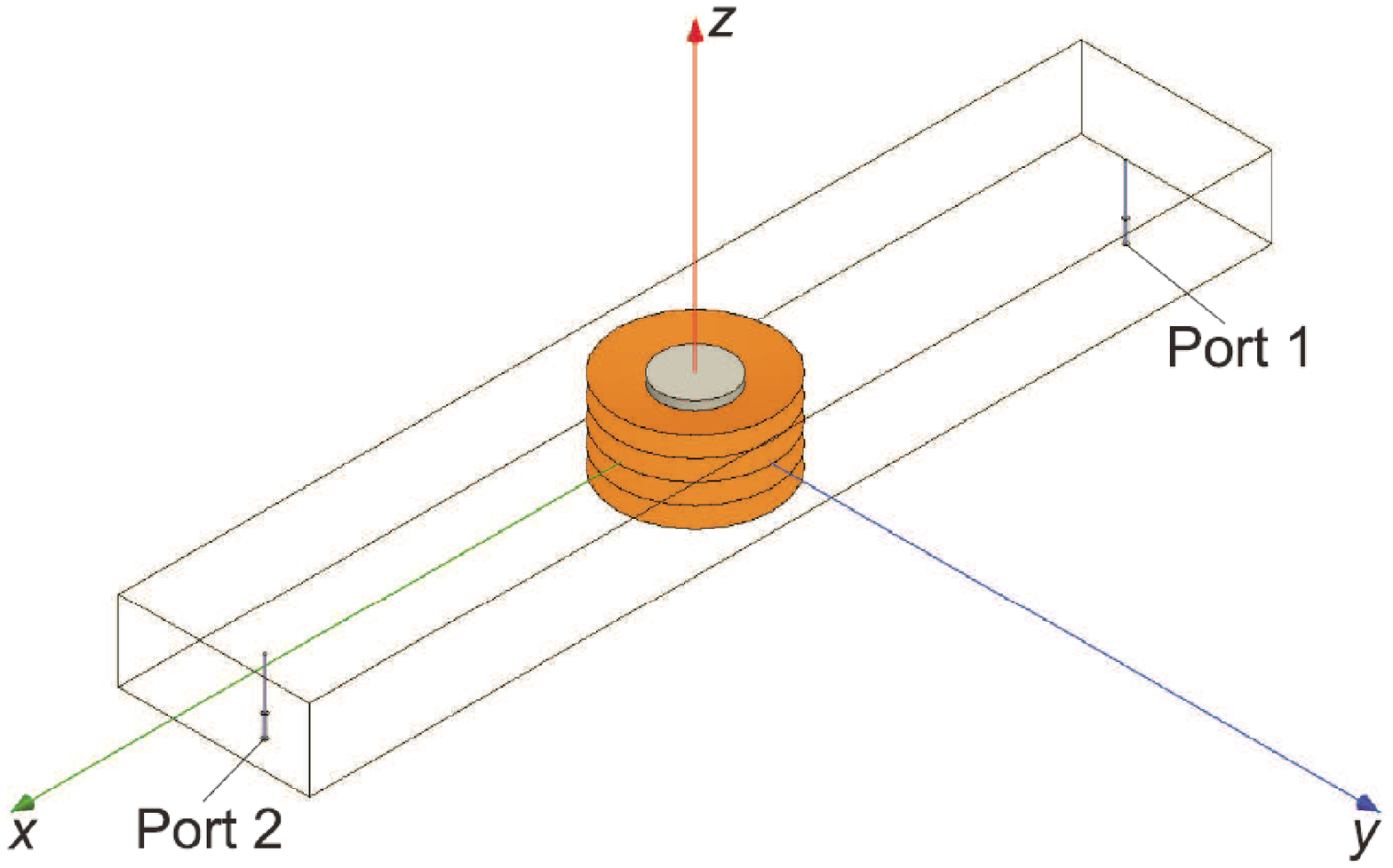, width=0.45\textwidth}}
\subfigure[]{\epsfig{file=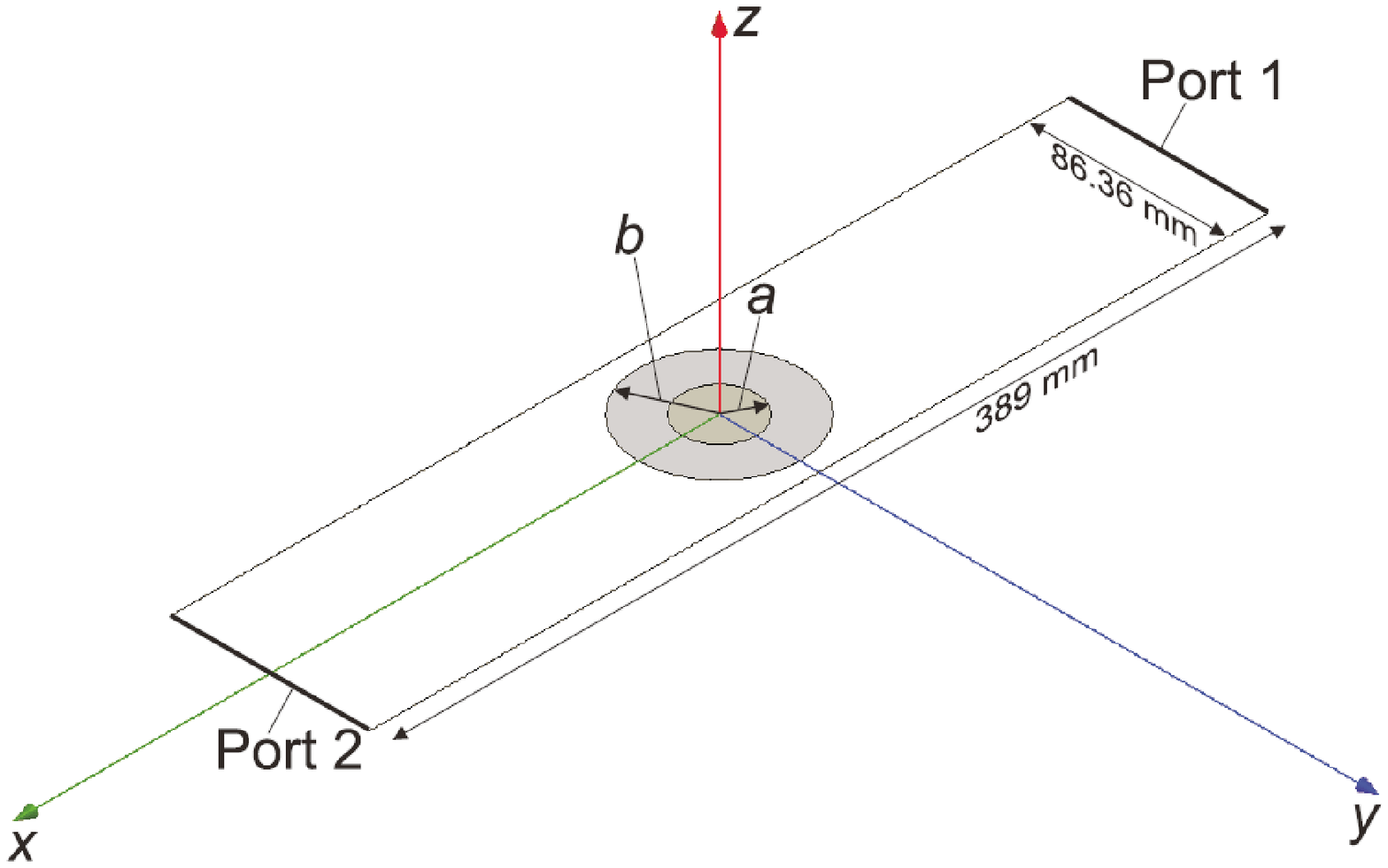, width=0.45\textwidth}}
\caption{A rectangular waveguide with a cloak and a cloaked object
inside. (a)~TL cloak -- the cloaked object is an array of metal
rods. (b)~MP cloak -- the cloaked object is a metal cylinder.
(c)~CT cloak -- the cloaked object is a metal cylinder. The
transmission data of structures in (a) and (b) are obtained
experimentally whereas the structure in (c) is simulated (the
simulation model is two-dimensional, i.e., the structure is
assumed to be infinite along the $z$-axis).} \label{cloaks}
\end{figure}

The coordinate-transformation cloak is formed of an anisotropic
material layer, the material parameters of which are designed with
the well-known cloak design equations that were used to design the
first experimental demonstration of a cloak of this
type.\cite{Schurig2006} To obtain the needed values for $\mu$ we
design the cloak to have a radially dependent component of the
permeability ($\mu_{\rho}$) having the optimal values at the
design frequency of 3~GHz. See Fig.~\ref{cloaks}c for an
illustration of the simulated structure. Suitable material
parameters at the design frequency of 3~GHz are found to
be\cite{Schurig2006}

\begin{equation}
\epsilon_z = \left(\frac{b}{b-a}\right)^2
-j0.002, \l{epsilon_z}
\end{equation}

\begin{equation}
\mu_\varphi = 1, \l{mu_phi}
\end{equation}

\begin{equation}
\mu_{\rho}(\rho) = \left(\frac{\rho-a}{\rho}\right)^2 - j0.006,
\l{mu_rho_c}
\end{equation} where $\rho$ is the distance from the center of the cloaked object and $a$ and $b$ are the inner and outer radius of the cloak, respectively. The losses are modelled according to the experimental setup.\cite{Schurig2006}

In a realization of such a cloak device, the parameters
$\epsilon_z$ and $\mu_\varphi$ are constant, but the radial
component of the permeability $\mu_\rho$ must be dispersive
(otherwise such values of $\mu_\rho$ cannot be realized with a
passive medium). Here we assume that $\mu_\rho$ has the Lorentzian
dispersion:

\begin{equation}
\mu_\rho(\rho,f) = 1- \frac{f_p^2}{f^2-f_0^2-jff_d},
\end{equation} where $f_p$ is the effective plasma frequency, $f_0$ the
resonance frequency, and $f_d$ the damping frequency.

This is a dispersion model typical for an artificial permeability
realized with split-ring-resonators. To enable good operation of
the cloak and mitigate the effect of dispersion, we position the
resonance frequency $f_0$ at a much lower frequency than the
operation frequency of the cloak. Note that this may not be
possible in practical designs due to too large inclusion
dimensions.

To satisfy the cloaking condition at the operating frequency of
$f_c=3$~GHz, we demand according to~\r{mu_rho_c} that

\begin{equation}
\mu_\rho(\rho,f_c) = \mu_{\rho,f_c}'-j\mu_{\rho,f_c}'' =
\left(\frac{\rho-a}{\rho}\right)^2 -j0.006,
\end{equation} from which it follows

\begin{equation}
f_d =
\frac{\mu_{\rho,f_c}''}{(1-\mu_{\rho,f_c}')f_c}(f_c^2-f_0^2),
\end{equation}

\begin{equation}
f_p^2 = \frac{\mu_{\rho,f_c}''}{f_c f_d}\left[(f_c^2-f_0^2)^2 +
f_c^2f_d^2\right]. \l{f_p}
\end{equation}

\begin{figure} [h!]
{\epsfig{file=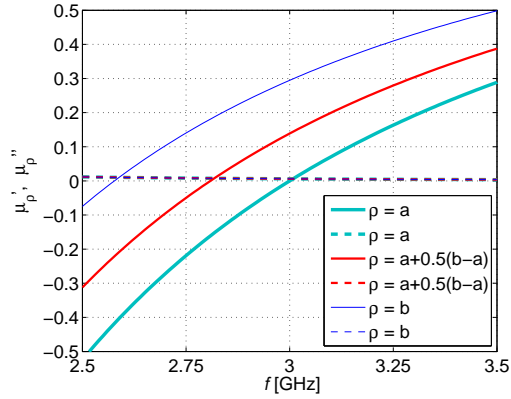, width=0.45\textwidth}} \caption{Frequency
dependence of $\mu_{\rho}$. Solid lines: real part
($\mu_{\rho}'$), dashed lines: imaginary part ($\mu_{\rho}''$).}
\label{CTcloak_parameters}
\end{figure}

To have a fair comparison with the other cloaks, here we use the
same inner and outer dimensions as for the MP cloak, i.e.,
$a=16$~mm (inner radius of the cloak) and $b=35$~mm (outer radius
of the cloak). The diameter of the cloaked perfectly conducting
cylinder is also the same as for the MP cloak, i.e., 30~mm. With
$f_c=3$~GHz and $f_0=1$~GHz and using~\r{mu_rho_c}--\r{f_p} we can
find the values for the function $\mu_{\rho}(\rho,f)$ and assign
these values to be continuous in the simulation model of
Fig.~\ref{cloaks}c. See Fig.~\ref{CTcloak_parameters} for the
values of the real and imaginary parts of $\mu_{\rho}$ as a
function of the frequency for three different values of $\rho$
inside the cloak. According to~\r{epsilon_z} $\varepsilon_z\approx
3.39 -j0.002$.


\section{Numerical verification of cloaking with the studied devices}

Before studying pulse propagation with the waveguide and cloak
systems shown  in Fig.~\ref{cloaks}, let us compare the cloaking
capabilities of the studied devices using frequency domain
simulations where the cloaks are placed in free space. We take the
cloaks (and cloaked objects) described in
Section~\ref{studied_cloaks} and model them numerically in free
space. The cloaks are modelled as being infinitely periodic along
the vertical ($z$) direction. By illuminating the cloaks with
plane waves having the electric field parallel to the $z$-axis we
can compute the total scattering cross sections of the cloaked and
uncloaked objects.\cite{Alitalo_TLcloakAPL}

\begin{figure} [t!]
{\epsfig{file=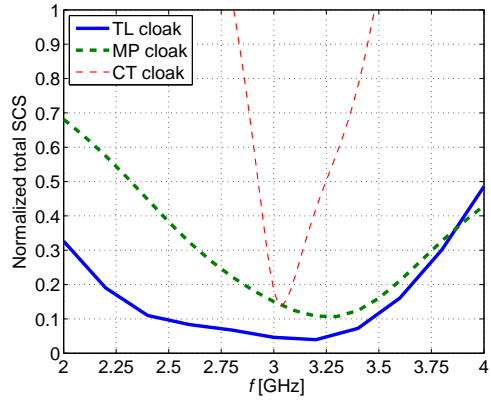, width=0.45\textwidth}} \caption{Simulated
total SCS of cloaked objects, normalized to the total SCS of
uncloaked objects. The results for the TL
cloak\cite{Alitalo_TLcloakAPL} and the MP
cloak\cite{Tretyakov_PPWGcloakPRL} are obtained with Ansoft
HFSS\cite{Ansoft} whereas the results for the CT cloak are
obtained with COMSOL Multiphysics.\cite{Comsol}} \label{SCStot}
\end{figure}

The results for the TL cloak and the MP cloak have been already
reported\cite{Alitalo_TLcloakAPL,Tretyakov_PPWGcloakPRL} and here
we compare these with the results for the CT cloak. From
Fig.~\ref{SCStot} it is obvious that the CT cloak has a much
narrower bandwidth as compared to the two other cloaks. This is an
expected consequence of the dispersive nature of the permeability
in the CT cloak, even though the resonance frequency of the
permeability was chosen to be far away from the design frequency
of 3~GHz. The minimum of the normalized total scattering cross
section (SCS) occurs at 3.04~GHz for the CT cloak.


\section{Measured and simulated transmission data}
\label{data}

\begin{figure} [h!]
\subfigure[]{\epsfig{file=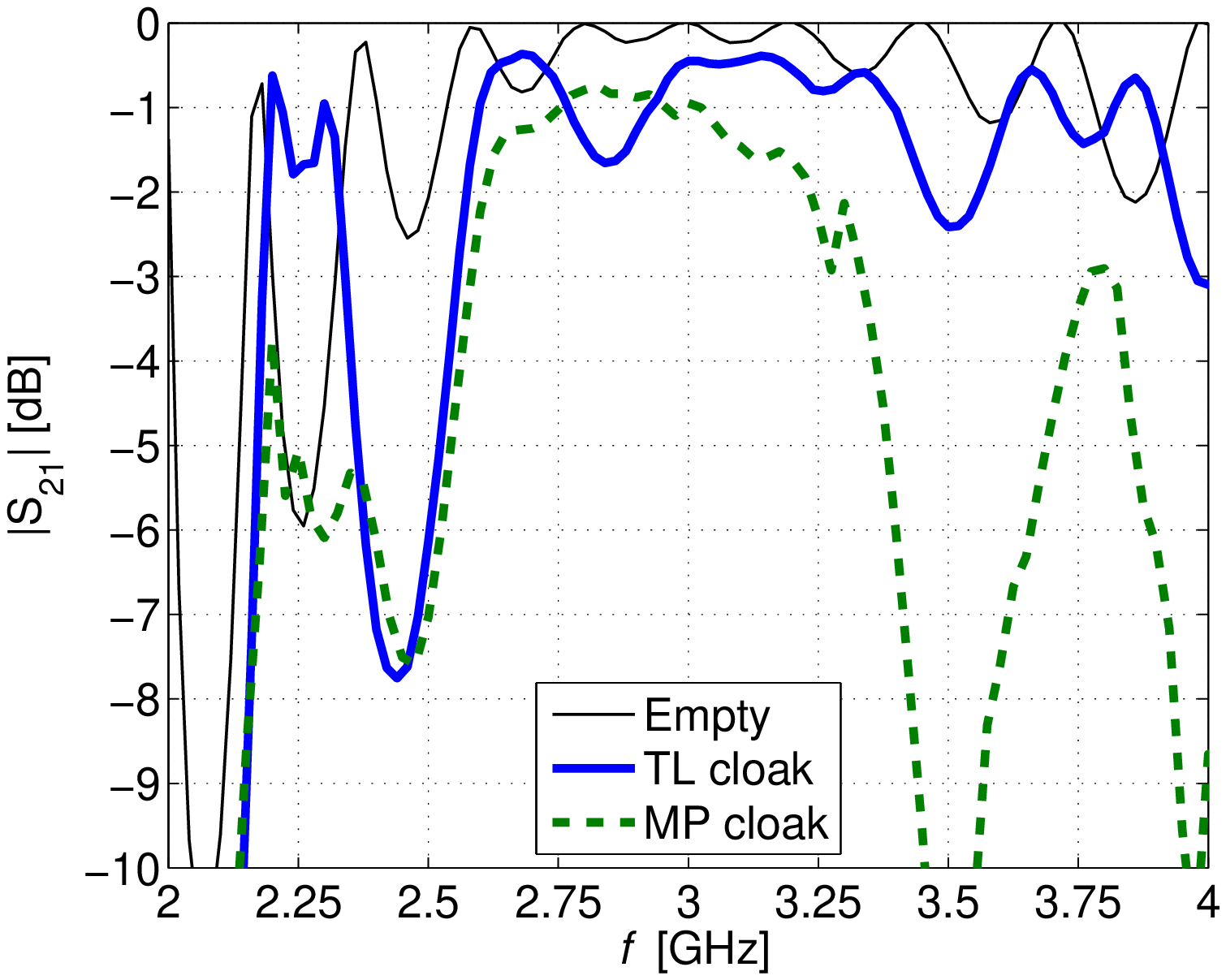, width=0.45\textwidth}}
\subfigure[]{\epsfig{file=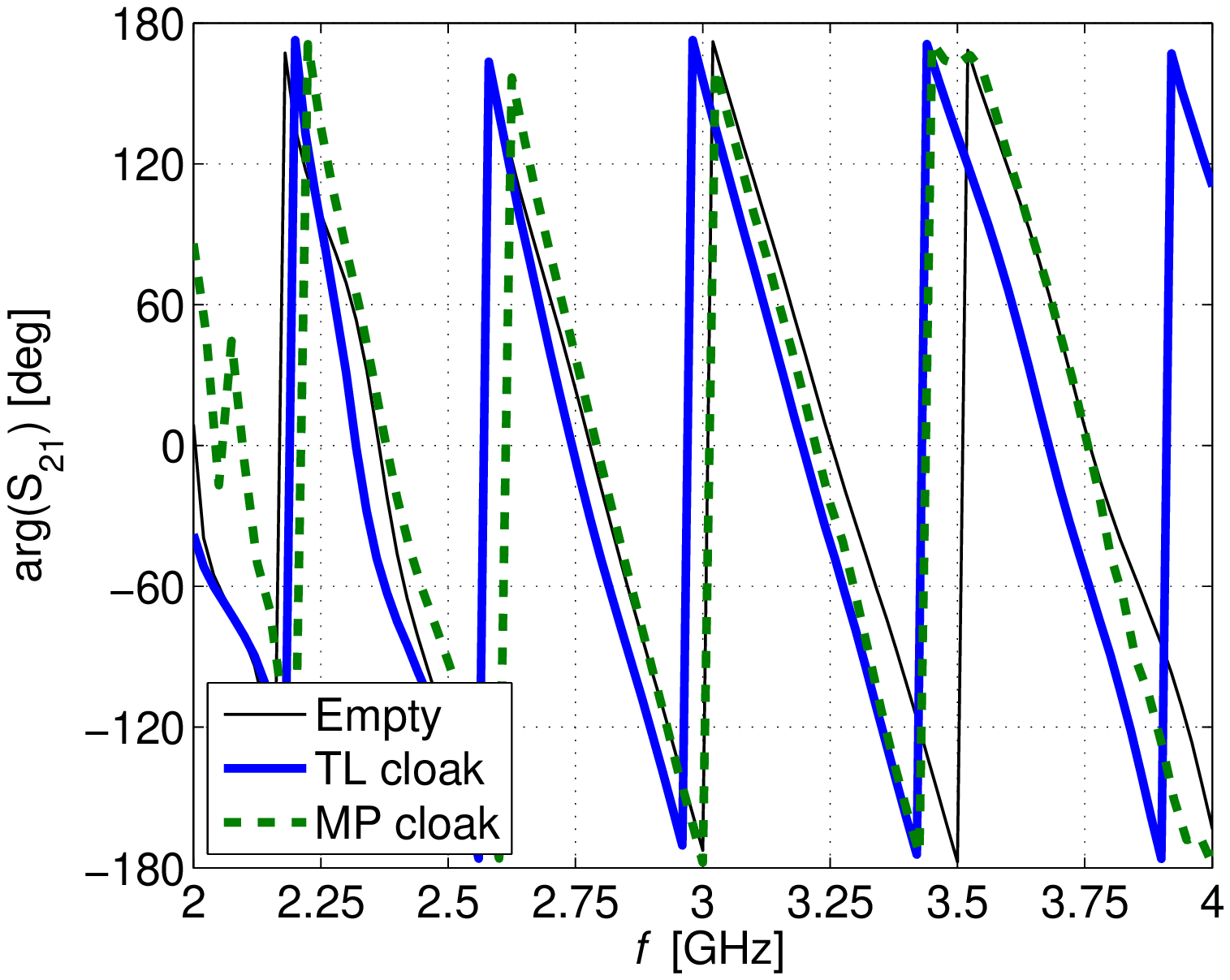, width=0.45\textwidth}}
\caption{(a)~Magnitude and (b)~phase of the transmission
coefficient. Measured
results.\cite{Alitalo_TLcloakAPL,Alitalo_TLcloakAPS2009,Tretyakov_PPWGcloakPRL}}
\label{s21_meas}
\end{figure}

\begin{figure} [h!]
\subfigure[]{\epsfig{file=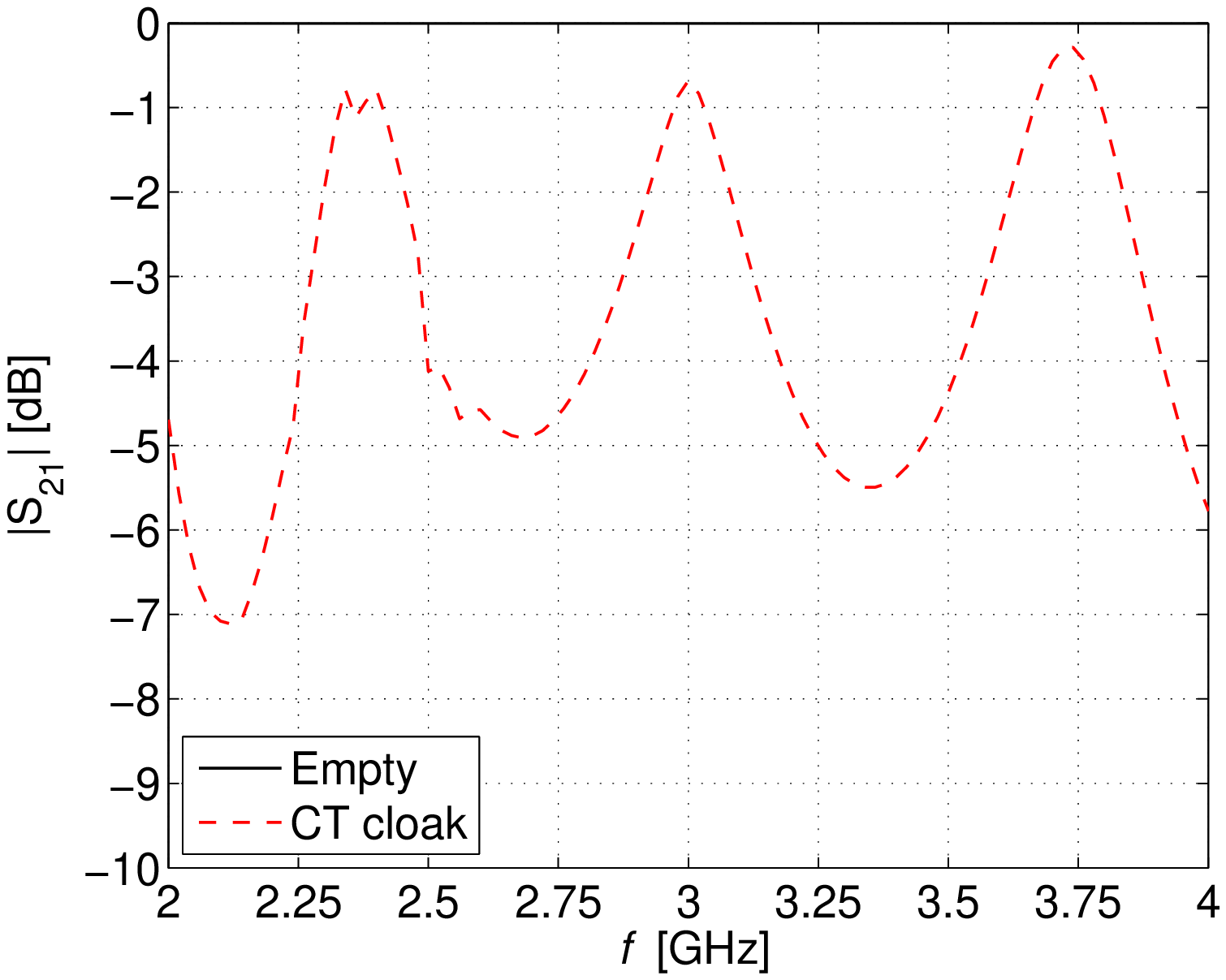, width=0.45\textwidth}}
\subfigure[]{\epsfig{file=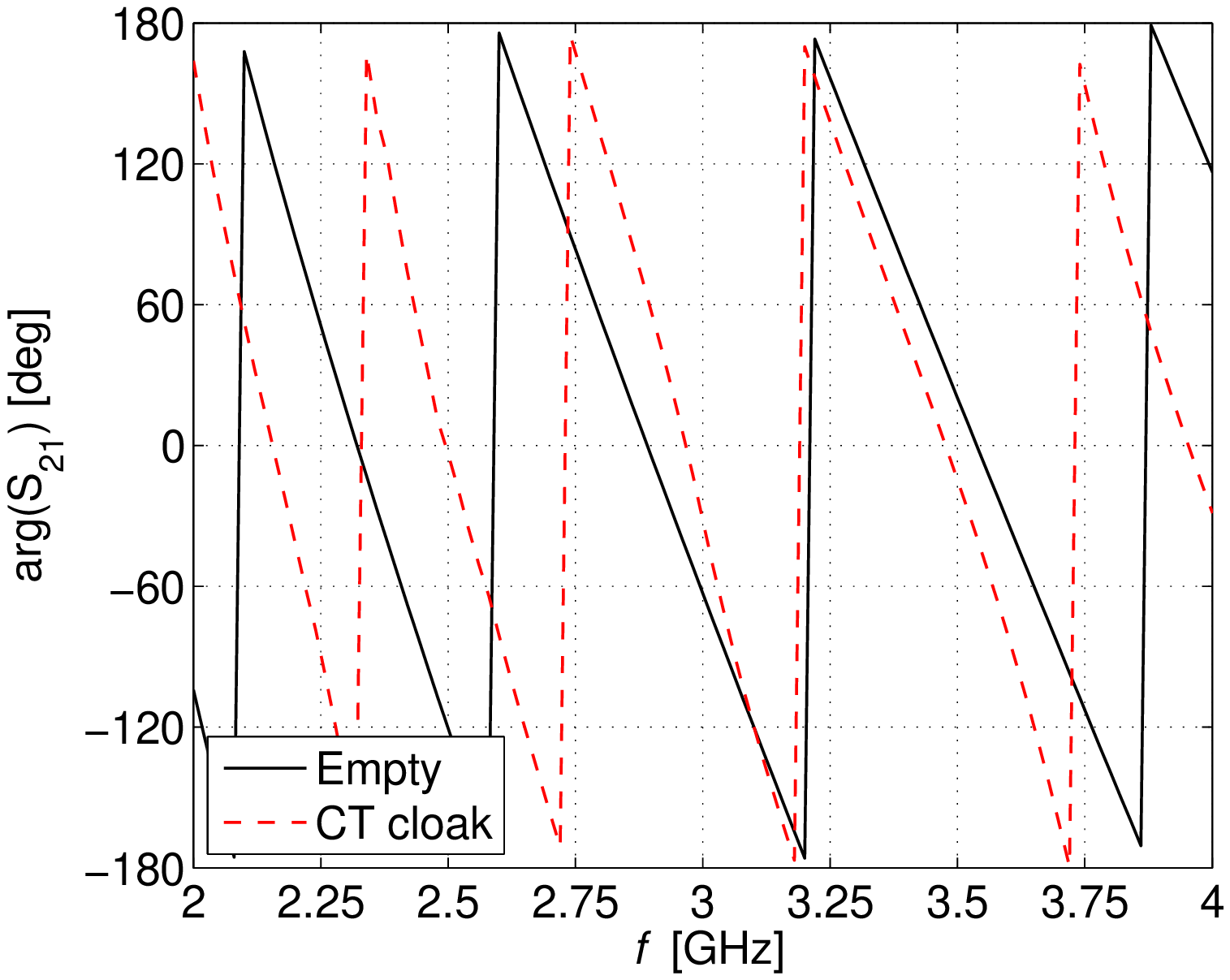, width=0.45\textwidth}}
\caption{(a)~Magnitude and (b)~phase of the transmission
coefficient. Simulated results. For the empty waveguide the
magnitude is 0~dB in this frequency range (above cut-off waveguide
with infinite height).} \label{s21_sim}
\end{figure}

The experiments and the transmission measurements for the empty
waveguide and the TL and MP cloaks shown in Fig.~\ref{cloaks}a,b
have been already
reported,\cite{Alitalo_TLcloakAPL,Alitalo_TLcloakAPS2009,Tretyakov_PPWGcloakPRL}
demonstrating good transmission properties around 3~GHz. Here we
compare these experimental results with the numerical results
obtained for the CT cloak of the same dimension. The measured
transmission magnitude and phase for the TL cloak and the MP cloak
are presented in Fig.~\ref{s21_meas} and the simulated
transmission magnitude and phase for the CT cloak are presented in
Fig.~\ref{s21_sim}. The simulations of the waveguide with the CT
cloak inside are conducted with COMSOL Multiphysics.\cite{Comsol}

%

The transmission magnitude for the CT cloak has several maxima.
The maximum at 3~GHz relates to the designed optimal cloaking
frequency, and the other maxima result from resonances inside the
cloak. From the simulated field distributions (e.g., animating the
time-harmonic electric field as a function of the phase) this is
obvious since at 3~GHz the electromagnetic wave that hits the
cloak travels inside the cloak, whereas at the frequencies
corresponding to the other maxima, the field oscillates inside the
cloak. While the transmitted amplitude is close to 0~dB at the
frequencies corresponding to these resonant maxima, the phase of
the transmitted wave strongly deviates from the empty-waveguide
case as shown in Fig.~\ref{s21_sim}b.

The transmission phase difference between the cloaked and empty
cases for various cloaks can be compared. For the TL cloak the
phase lags as compared to the empty waveguide and this lag is
growing with increasing frequency. This is expected since the
wavenumber of waves travelling inside the TL cloak is not equal to
the free-space wavenumber.\cite{Alitalo_review,Alitalo_TLcloakTAP}
In contrast, for the MP cloak at around 3~GHz the transmission
phases for cloaked and empty cases are almost identical. Below and
above this frequency the phase in the MP cloak has a small lead
and lag, respectively.

For the CT cloak we would expect to have identical phase shift
for cloaked and empty cases at the design frequency of 3~GHz.
However, the matching of the transmission phase occurs at
approximately 3.1~GHz. This can be a result of the simplification
of the cloak's material parameters\cite{Schurig2006} or the
influence of losses.

When comparing Figs.~\ref{s21_meas}b and~\ref{s21_sim}b it should
noted that the electrical lengths from port~1 to port~2 are
different in the two figures. This is because the measured results
include the feed probes whereas the simulated results do not, see
Fig.~\ref{cloaks}.

\section{Pulse distortion}

\begin{figure} [h!]
\subfigure[]{\epsfig{file=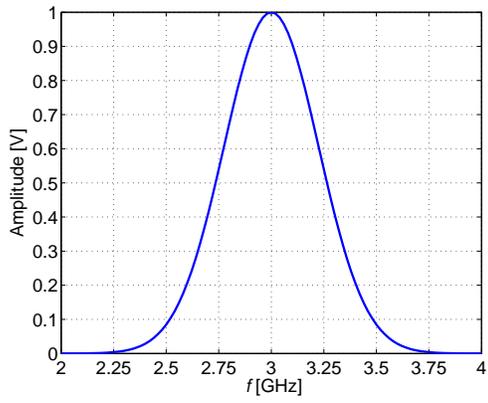, width=0.45\textwidth}}
\subfigure[]{\epsfig{file=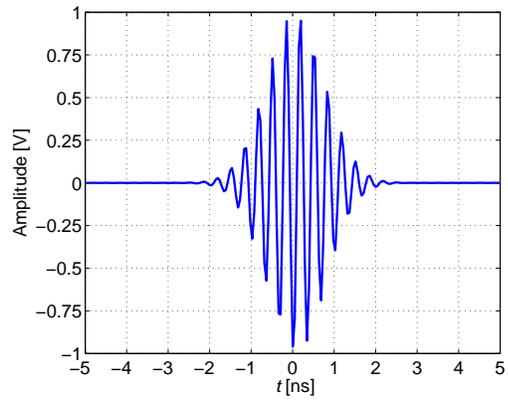, width=0.45\textwidth}}
\caption{Magnitude of the pulse at the input port (port~1) of the
waveguide. (a)~Frequency domain. (b)~Time domain.}
\label{orig_pulse}
\end{figure}

Next we study pulse propagation inside the
waveguides enclosing the different cloaks. This can be
investigated analytically by using the frequency-domain transmission data.
First we choose the center frequency of a Gaussian
pulse ($f_{c,G}$) that we want to transmit through the waveguide and
apply a Gaussian filter to create a time-domain signal $y(t)$

\begin{equation}
y(t)=\sin(2\pi f_{c,G}t)e^{-\frac{t^2}{2\sigma^2}},
\end{equation} where $\sigma$ is the standard deviation and the
center of the pulse corresponds to time $t=0$. For $f_{c,G}=3$~GHz
and $\sigma=7.07\cdot 10^{-10}$ the pulse looks in the frequency
and time domains as shown in Fig.~\ref{orig_pulse}.

\begin{figure} [h!]
\subfigure[]{\epsfig{file=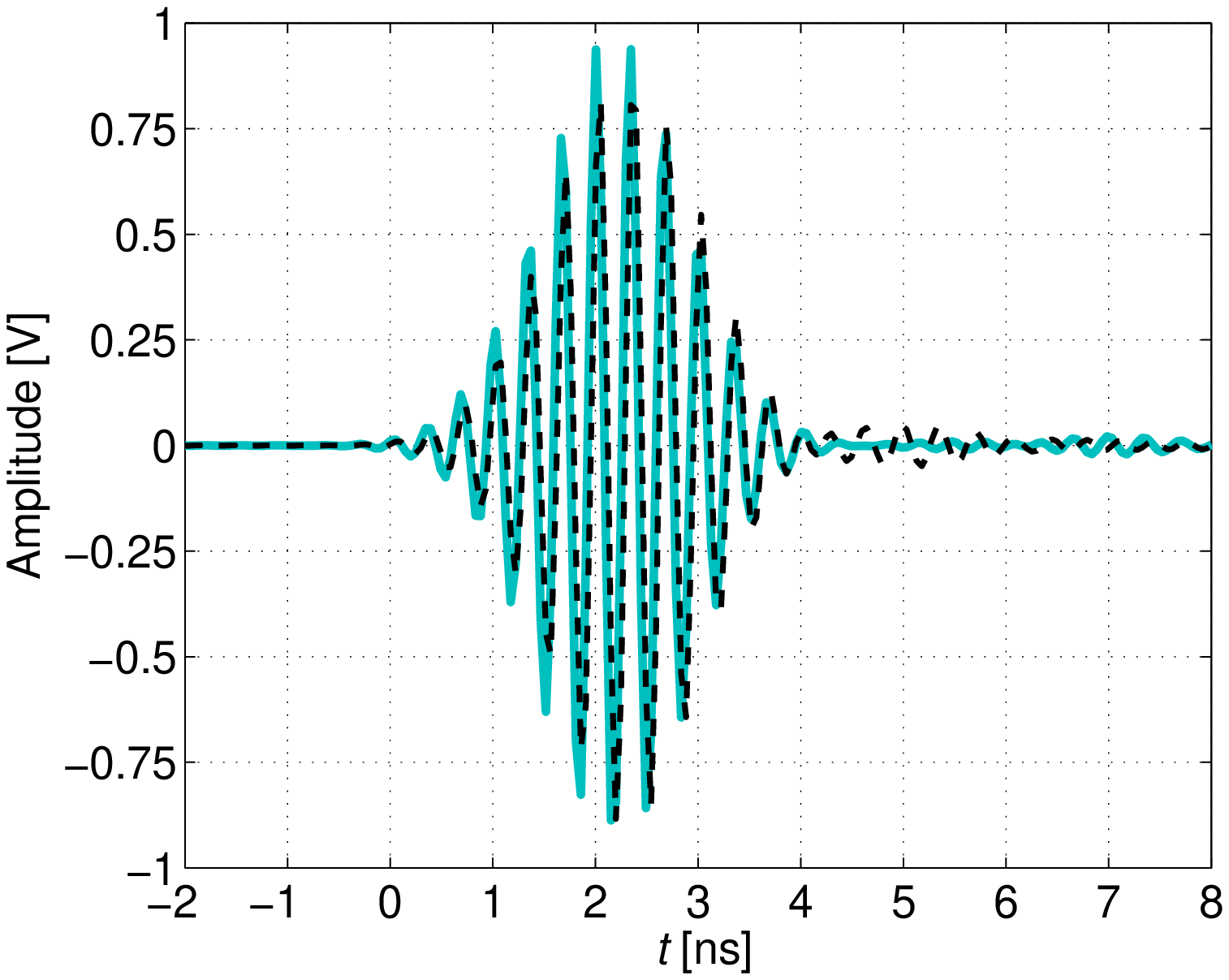, width=0.42\textwidth}}
\subfigure[]{\epsfig{file=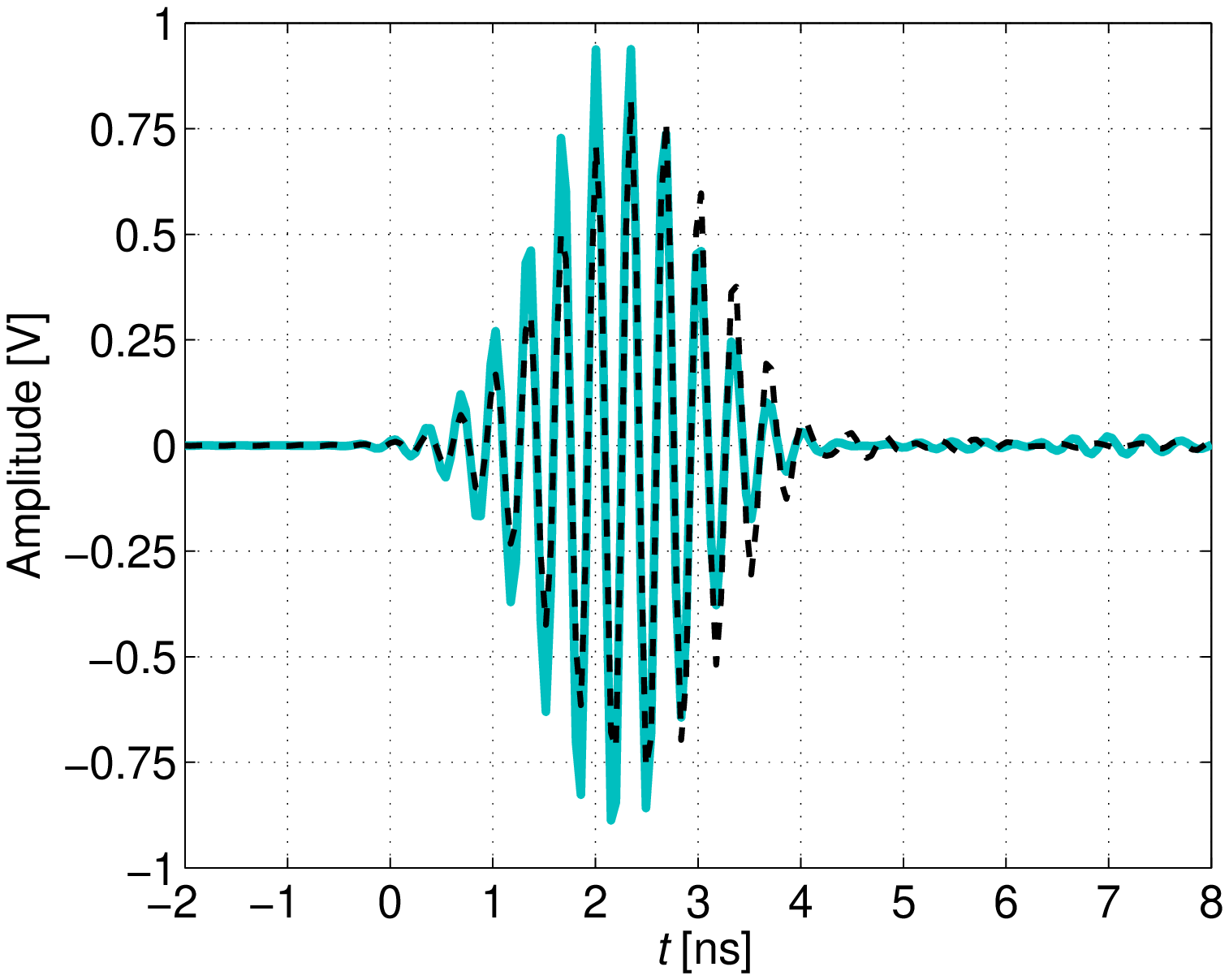, width=0.42\textwidth}}
\subfigure[]{\epsfig{file=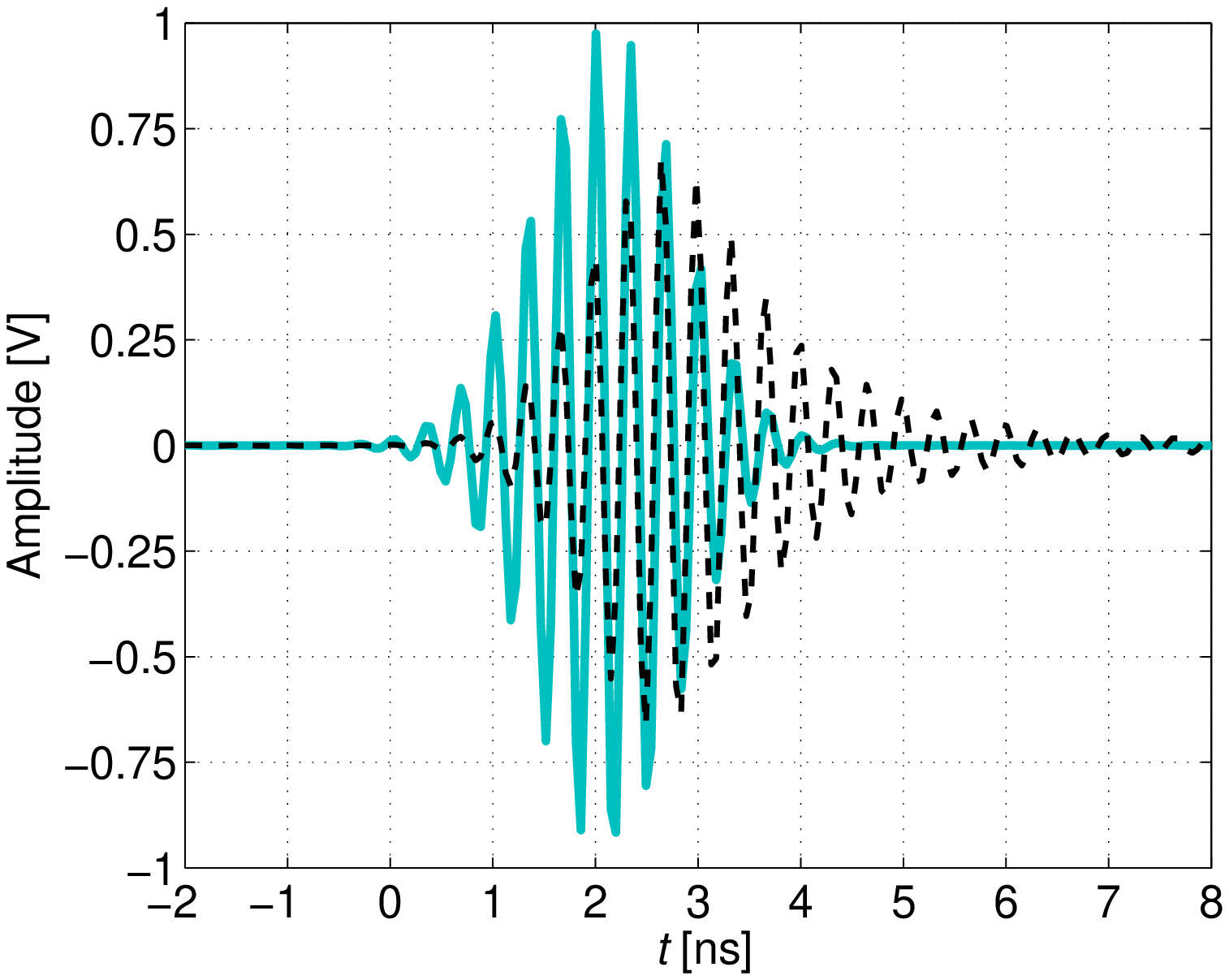, width=0.42\textwidth}}
\caption{Pulse shape in time domain at the output port of the
waveguide. $f_{c,G}=3$~GHz and $\sigma=7.07\cdot 10^{-10}$. Solid
lines: empty waveguide, dashed lines: waveguide with a cloak
inside. (a)~TL cloak (measured transmission data used). (b)~MP
cloak (measured transmission data used). (c)~CT cloak (simulated
transmission data used).} \label{pulses}
\end{figure}

We apply the Fourier transform to the time-domain pulse and
multiply each frequency component by the complex transmission
coefficients presented in Section~\ref{data}. Then by applying the
inverse Fourier transform we find the time-domain signal at the
output of the waveguide section (at port~2). Fig.~\ref{pulses}
presents these time-domain signals for all studied cloaks together
with the signals at the output of an empty waveguide when using
the input pulse of Fig.~\ref{orig_pulse}.

From Fig.~\ref{pulses} we can conclude that for the TL cloak and
the MP cloak the pulse shape reasonably well corresponds to the
pulse at the output port of an empty waveguide. For the CT cloak,
however, the pulse is clearly distorted and there is a
considerable ``tail'' because the energy inside the cloak travels
slower than in free space (the slower the closer to the inner part
of the cloak\cite{Alitalo_review,Chen}).

\begin{figure} [h!]
{\epsfig{file=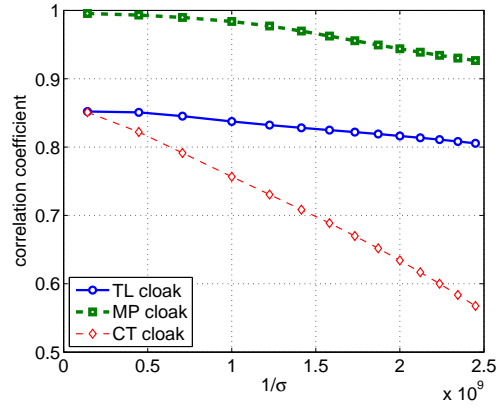, width=0.45\textwidth}}
\caption{Correlation coefficients of the signals transmitted
through cloaks and the signal through the empty waveguide.
$f_{c,G}=3$~GHz.} \label{corr1}
\end{figure}

From the time-domain results in Fig.~\ref{pulses} it is difficult
to define a numerical measure on how well or bad the cloaks work.
Therefore, we calculate the correlation coefficients between the
pulse which goes through the empty waveguide and the pulses which
go through the cloaks. See Fig.~\ref{corr1} for the results as a
function of the inverse of the standard deviation while
$f_{c,G}=3$~GHz is kept constant. Small values of $1/\sigma$
correspond to pulses which are wide in the time-domain and narrow
in the frequency domain, whereas large values of $1/\sigma$
correspond to pulses which are narrow in the time-domain and wide
in the frequency domain.

All the cloaks work well for pulses which are narrow in the
frequency domain. As the pulse bandwidth increases, the CT cloak's
operation deteriorates much faster than the operation of the other
cloaks. This is expected since the CT cloak has a narrow bandwidth
(see Fig.~\ref{SCStot}) and stronger dispersion as compared to the
other cloaks.

\begin{figure} [h!]
{\epsfig{file=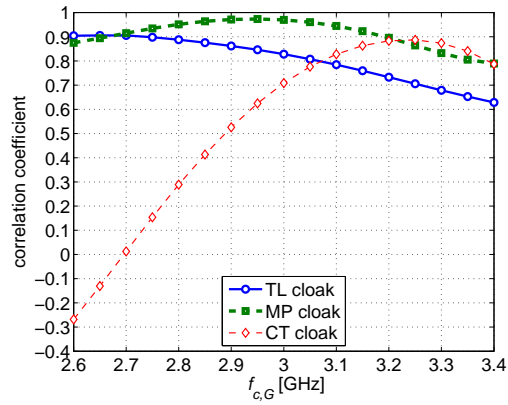, width=0.45\textwidth}}
\caption{Correlation coefficients of the signals transmitted
through cloaks and the signal through the empty waveguide.
$\sigma=7.07\cdot 10^{-10}$.} \label{corr2}
\end{figure}

Another interesting issue is to study how the choice of the center
frequency of the pulse ($f_{c,G}$) affects the results. Here we
use a constant standard deviation of $\sigma=7.07\cdot 10^{-10}$
and plot the correlation coefficients as functions of $f_{c,G}$,
see Fig.~\ref{corr2}. We see that the TL cloak works better for
pulses for which the center frequency is significantly lower than
the design frequency of 3~GHz while the MP cloak works best when
the center frequency is approximately 2.9~GHz. This can be
explained by Fig.~\ref{s21_meas}b which shows that the optimal
matching of the transmission phases of the empty waveguide and the
TL cloak occurs at around 2.6~GHz while for the MP cloak the
optimal frequency is approximately 2.9~GHz.

The CT cloak appears to work better for pulses with $f_{c,G}$ at
frequencies slightly higher than the design frequency of 3~GHz.
Again, this can be partly explained by the transmission phase
shown in Fig.~\ref{s21_sim}b according to which the optimal
matching of the transmission phases for the empty waveguide and
the CT cloak occurs at 3.1~GHz. Also the blueshift
effect\cite{Zhang1} may cause the cloak to work better for pulses
above the design frequency.


\section{Conclusions}

Three different cloak designs have been studied and their
operation compared with each other: The transmission-line cloak,
the metal-plate cloak, and the coordinate-transformation cloak.
The cloaking capabilities of these devices are first studied with
numerical frequency-domain simulations of the cloaks in free space
with plane waves illuminating the cloaks. This type of comparison
enables the study of the reduction of the total scattering cross
sections, but it does not give information about distortion of
time-domain pulses.

To compare the cloaks' operation for time-domain signals we use
measured and simulated data for the transmission coefficients of
rectangular waveguides with the cloaks (and cloaked objects)
inside. With this transmission data we can analytically study how
time-domain signals travel through the waveguides. It is shown
that the transmission-line cloak and the metal-plate cloak work
better than the coordinate-transformation cloak for pulses which
are narrow in the time-domain. For pulses which are wide in
time-domain (or, in other words, which are narrow in frequency
domain) all the cloaks offer reasonably good performance.

\section*{Acknowledgement}

\noindent
This work has been partially funded by the Academy of
Finland through the Center-of-Excellence program.


\clearpage

\section*{Figures}

\end{document}